\documentclass[a4paper,11pt]{article}
\usepackage{pos}
\usepackage{graphicx}
\usepackage[caption=false]{subfig}
\usepackage{float}
\usepackage{wrapfig}
\usepackage{multirow}
\title{Interpretability of an Interaction Network for identifying $H \rightarrow b\bar{b}$ jets}

\author*[a]{Avik Roy}
\author[a]{Mark S. Neubauer}

\affiliation[a]{Department of Physics, University of Illinois at Urbana-Champaign \\
1110 W Green St Loomis Laboratory, Urbana, Illinois 61801, United States of America}

\emailAdd{avroy@illinois.edu}
\emailAdd{msn@illinois.edu}

\abstract{
Multivariate techniques and machine learning models have found numerous applications in High Energy Physics (HEP) research over many years. In recent times, AI models based on deep neural networks are becoming increasingly popular for many of these applications. However, neural networks are regarded as black boxes- because of their high degree of complexity it is often quite difficult to quantitatively explain the output of a neural network by establishing a tractable input-output relationship and information propagation through the deep network layers. As explainable AI (xAI) methods are becoming more popular in recent years, we explore interpretability of AI models by examining an Interaction Network (IN) model designed to identify boosted $H\to b\bar{b}$ jets amid QCD background. We explore different quantitative methods to demonstrate how the classifier network makes its decision based on the inputs and how this information can be harnessed to reoptimize the model- making it simpler yet equally effective. We additionally illustrate the activity of hidden layers within the IN model as Neural Activation Pattern (NAP) diagrams. Our experiments suggest NAP diagrams reveal important information about how information is conveyed across the hidden layers of deep model. These insights can be useful to effective model reoptimization and hyperparameter tuning.
}

\FullConference{%
  41st International Conference on High Energy physics - ICHEP2022\\
  6-13 July, 2022\\
  Bologna, Italy
}

\begin{document}
\maketitle

\section{Introduction}

Owing to their intricate internal structure, neural networks (NNs) have often been treated as \textit{black boxes}. It is difficult to understand how different input features contribute to the network's computational process and how the inter-connected neural pathways convey information. In recent years, advances in \textit{explainabale} Artificial Intelligence (xAI)~\cite{MILLER20191} have made it possible to build intelligible relationship between an AI model's inputs, architecture, and predictions~\cite{xAI-intro}. 
xAI has been successful in learning the underlying physics of a number of problems in high energy detectors~\cite{9302535}. In this work, we apply state-of-the art xAI techniques in interpreting an Interaction Network (IN) model~\cite{IN} developed to identify boosted $H\to b\bar{b}$ jets from QCD background.

\section{Evaluating Feature Importance for the IN Model}

\begin{figure}[!h]
  \begin{minipage}[c]{0.65\columnwidth}
    \centering
    \includegraphics[width=\columnwidth]{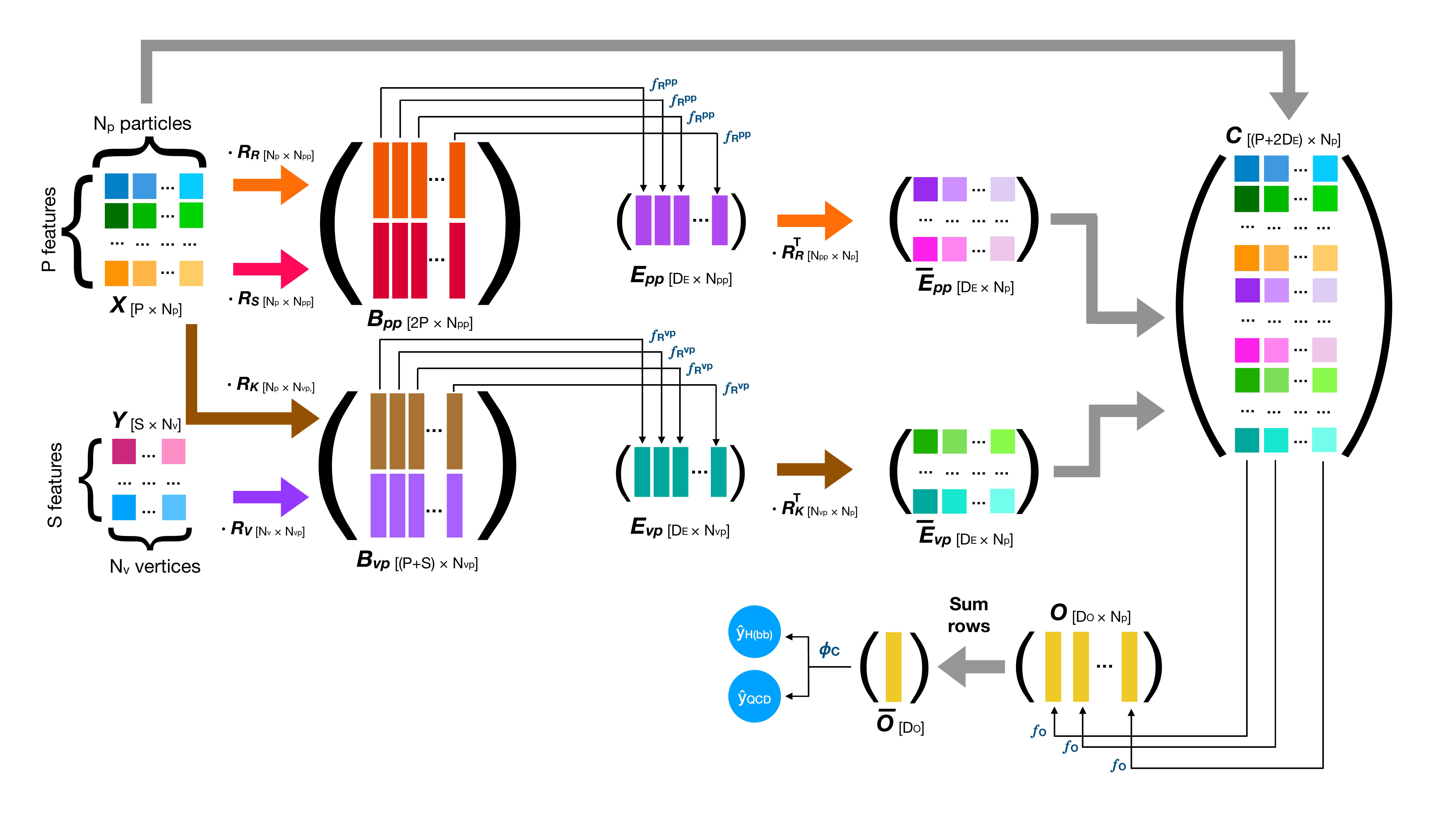}
  \end{minipage}%
  \begin{minipage}[c]{0.150\columnwidth}
    \centering
\begin{tabular}[b]{ |c|c|c| }
  \hline
  \multicolumn{2}{|c|}{Default IN hyperparameters} \\
  \hline \hline
  $(P, N_p, S, N_v)$ &  $(30, 60, 14, 5)$ \\ \hline
  Hidden layers &  3\\ \hline
 Hidden layer   &  \multirow{2}{*}{60}\\ 
 dimension & \\ \hline
  $(D_e, D_o)$ &  (20, 24)\\ \hline
  Activation &  ReLU\\ \hline%
\end{tabular}
\end{minipage}
\caption{A schematic diagram of the network architecture and dataflow in the IN model. This image is taken from Ref.~\cite{IN}. The choice of model hyperparameters and input data dimensions for the baseline model is given in the accompanying table.}
\label{fig:IN-arch}
\end{figure}

Figure~\ref{fig:IN-arch} shows the IN model architecture and tabulates the default hyperparameters and data dimensions. This network is built to train on graph data structure whose nodes comprise of $N_p$ particle tracks, each with $P$ features, and $N_v$ secondary vertices, each with $S$ features, associated with the jet. The physical description of each feature is given in Appendix C of ref.~\cite{IN}. It creates a fully connected directed graph  with $N_{pp} = N_p(N_p - 1)$ edges for the particle tracks. A separate graph with $N_{vp} = N_vN_p$ generates all possible connections between the particle tracks and the secondary vertices. 

The node level features for the track-track (track-vertex) graph are
transformed to edge level features via a couple of interaction matrices, identified as $R_{R[N_p \times N_{pp}]}$ and $R_{S[N_p \times N_{pp}]}$ 
($R_{K[N_p \times N_{vp}]}$ and $R_{V[N_v \times N_{vp}]}$). These edge-level features are transformed via fully connected NNs, respectively called $f_R^{pp}$ and $f_R^{vp}$, to obtain two $D_e$ dimensional internal state representation of these graphs. 

These internal state edge-level representations
are transferred back to the track-level representations by the interaction matrices. 
These particle-level internal state representations are concatenated with the original track features creating a $(P+2D_e) \times N_p$ dimensional feature space.
The trainable dense MLP $f_O$ creates the post-interaction $D_O$ dimensional internal representation that are summed over the tracks and then linearly combined to produce a two-dimensional output, which is transformed to individual class probabilities via a softmax function.  

\begin{wrapfigure}{L}{0.55\textwidth}
  \begin{center}
    \centering
    \includegraphics[width=0.53\textwidth]{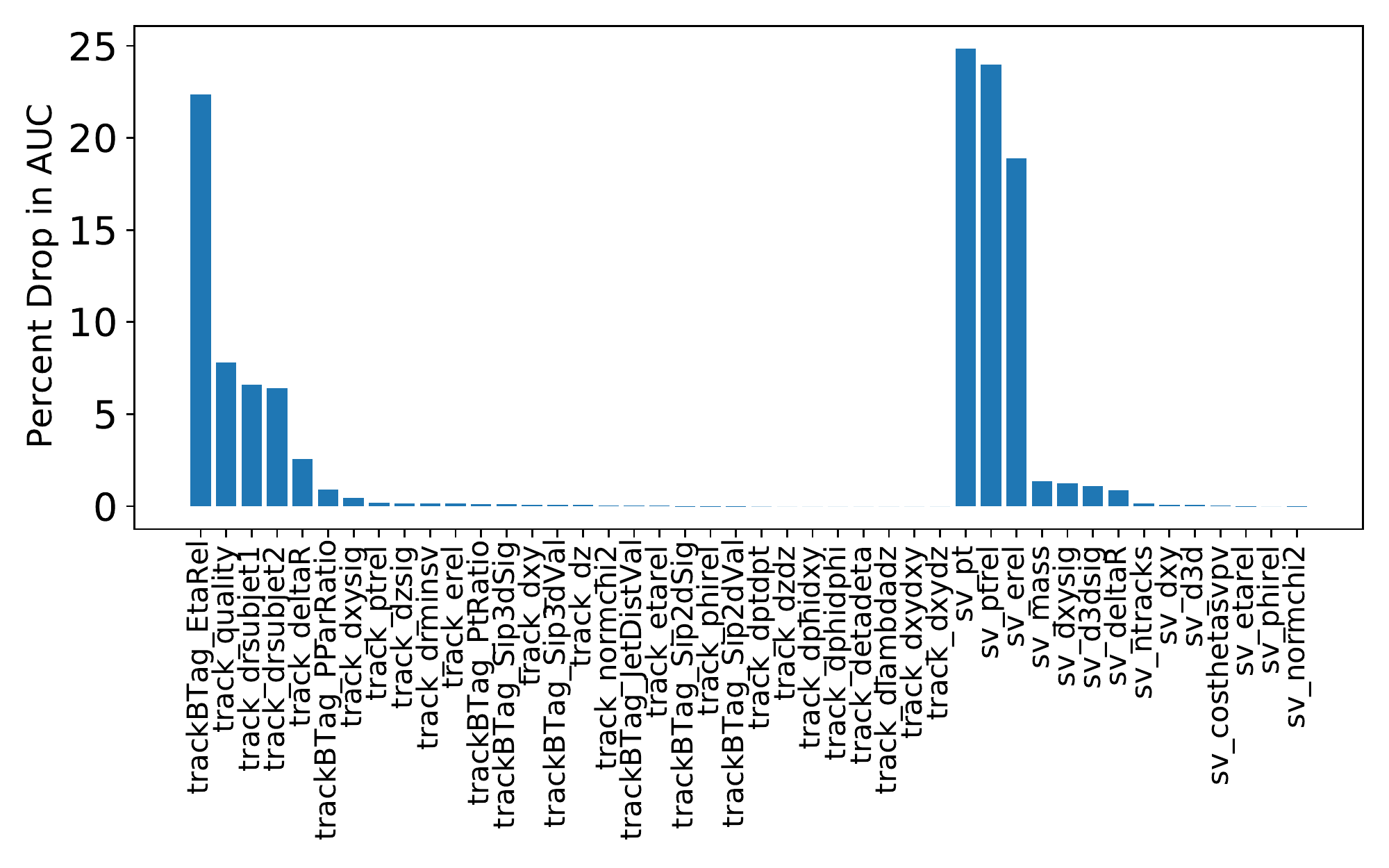}
  \end{center}%
\caption{Change in AUC score with respect to the baseline model when each of the  track and  secondary vertex features is individually masked during inference with the trained baseline model. }
\label{fig:dAUC}
\end{wrapfigure}

In order to realize which features play the most important role in the IN's decision making process, first we train the model with its default settings (the \textit{baseline} model). 
We mask one feature at a time for all input tracks or secondary vertices by replacing the corresponding entries by zero values. We obtain inference from the trained model and evaluate the Area Under the Curve (AUC) for Region Operator Characteristic (ROC) curve (ROC AUC score) from the model inference. The change observed in the AUC score for masking each of the features can be seen in Figure~\ref{fig:dAUC}. It shows that many of these input features have a rather small impact on the model's overall performance, reflected by the very small change in AUC score. 
We can inspect the importance of these features for individual tracks and vertices by the Layerwise Relevance Propagation (LRP) technique~\cite{LRP-NN, LRP-overview}. 
Since some of the input features show high degree of  correlation with each other, we use the LRP-$\gamma$ method described in ref.~\cite{LRP-overview}, which is designed to skew the LRP score distributions to nodes with positive weights in the network.
In order to apply the LRP method for the IN model, we defined custom propagation functions for 
(i) the aggregation of internal representation 
and (ii) the transformation via interaction matrices. 

\begin{figure}[h]
\centering
\subfloat[]{
\includegraphics[width=0.5\textwidth]{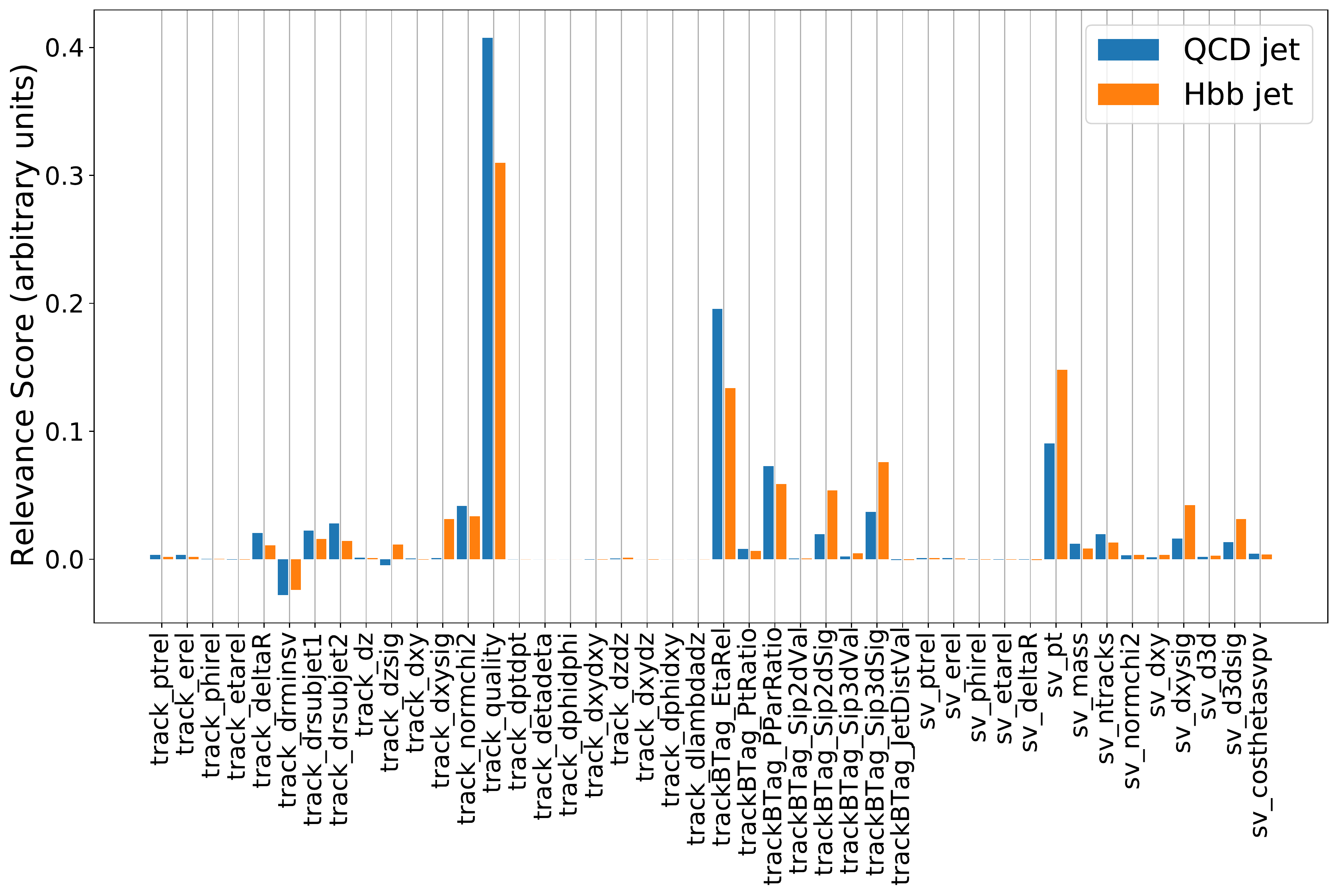}
\label{fig:LRP-feats}            
}
\subfloat[]{
\includegraphics[width=0.5\textwidth]{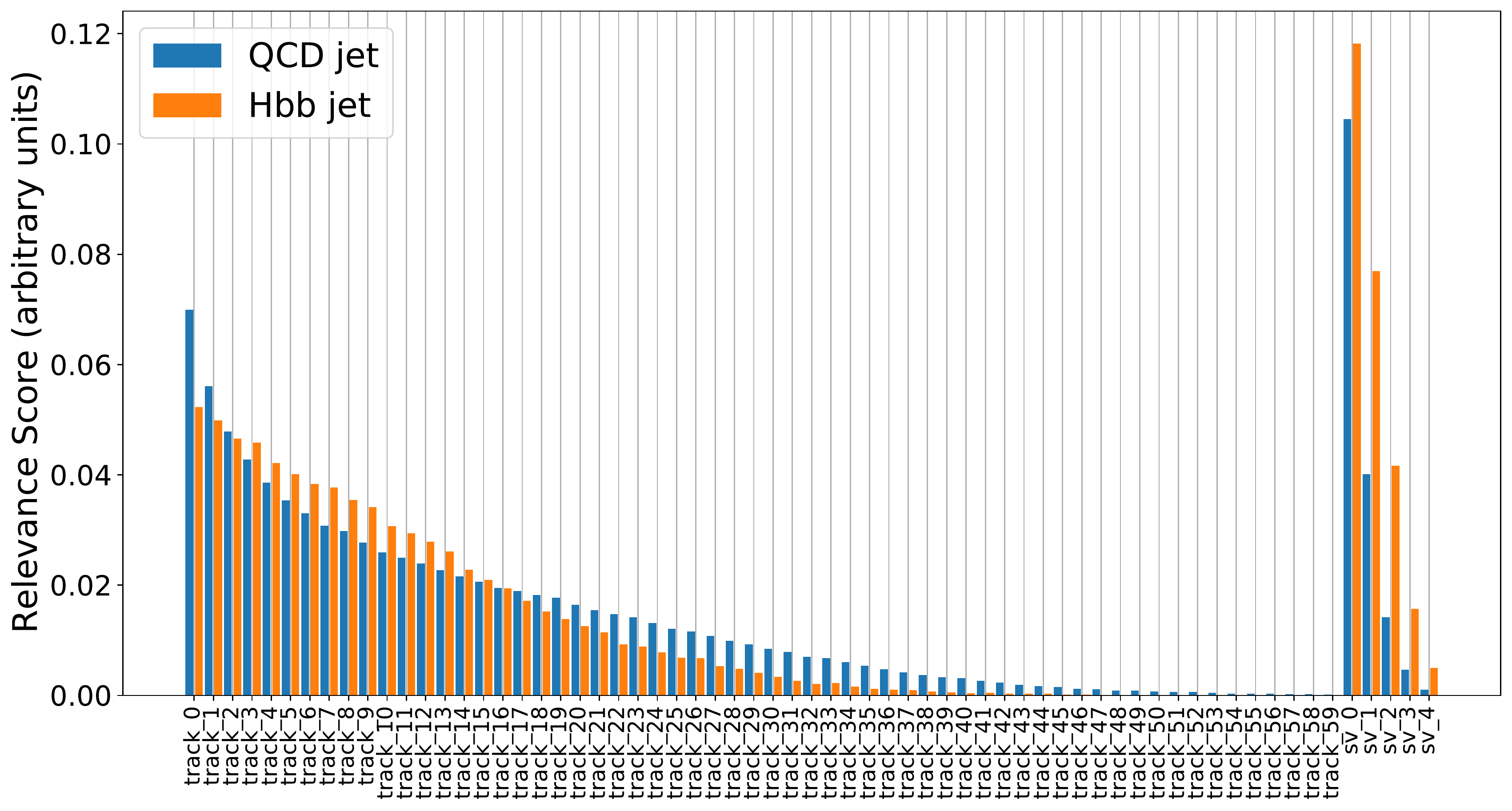}
\label{fig:LRP-trackNverts}            
}
\caption{Average relevance scores attributed to \protect\subref{fig:LRP-feats} input track and secondary vertex features and \protect\subref{fig:LRP-trackNverts} individual tracks and secondary vertices. The tracks and secondary vertices are ordered according to their relative energy with respect to the jet energy.}
\label{fig:LRP}
\end{figure}

We show the average scores attributed to the different features for QCD and $H \rightarrow b\bar{b}$ jets in figure~\ref{fig:LRP-feats}. When compared with the change in AUC score by individual features in figure~\ref{fig:dAUC}, the  features with largest relevance scores coincide with the features that individually cause the largest drop in AUC score. We additionally observe that vertices features play a more important role in identifying the $H \rightarrow b\bar{b}$ jets. This behavior is also justified from a physics stand point, since the presence of high energy secondary vertices is an important signature for jets from $b$ quark because of its relatively longer lifetime. 

\begin{figure}[H]
\centering
\subfloat[]{
\includegraphics[width=0.33\textwidth]{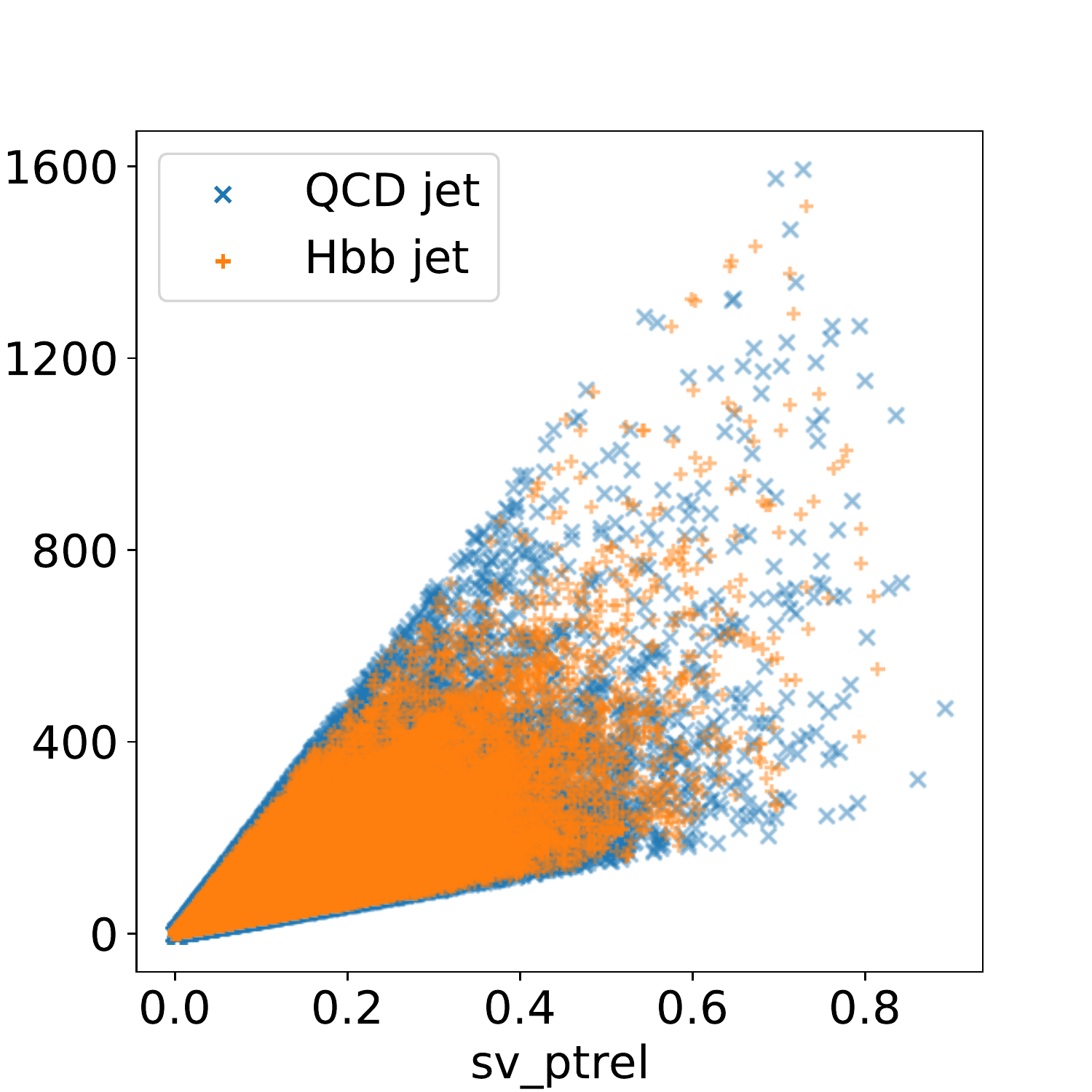}
\label{fig:svptrel}            
}
\subfloat[]{
\includegraphics[width=0.30\textwidth]{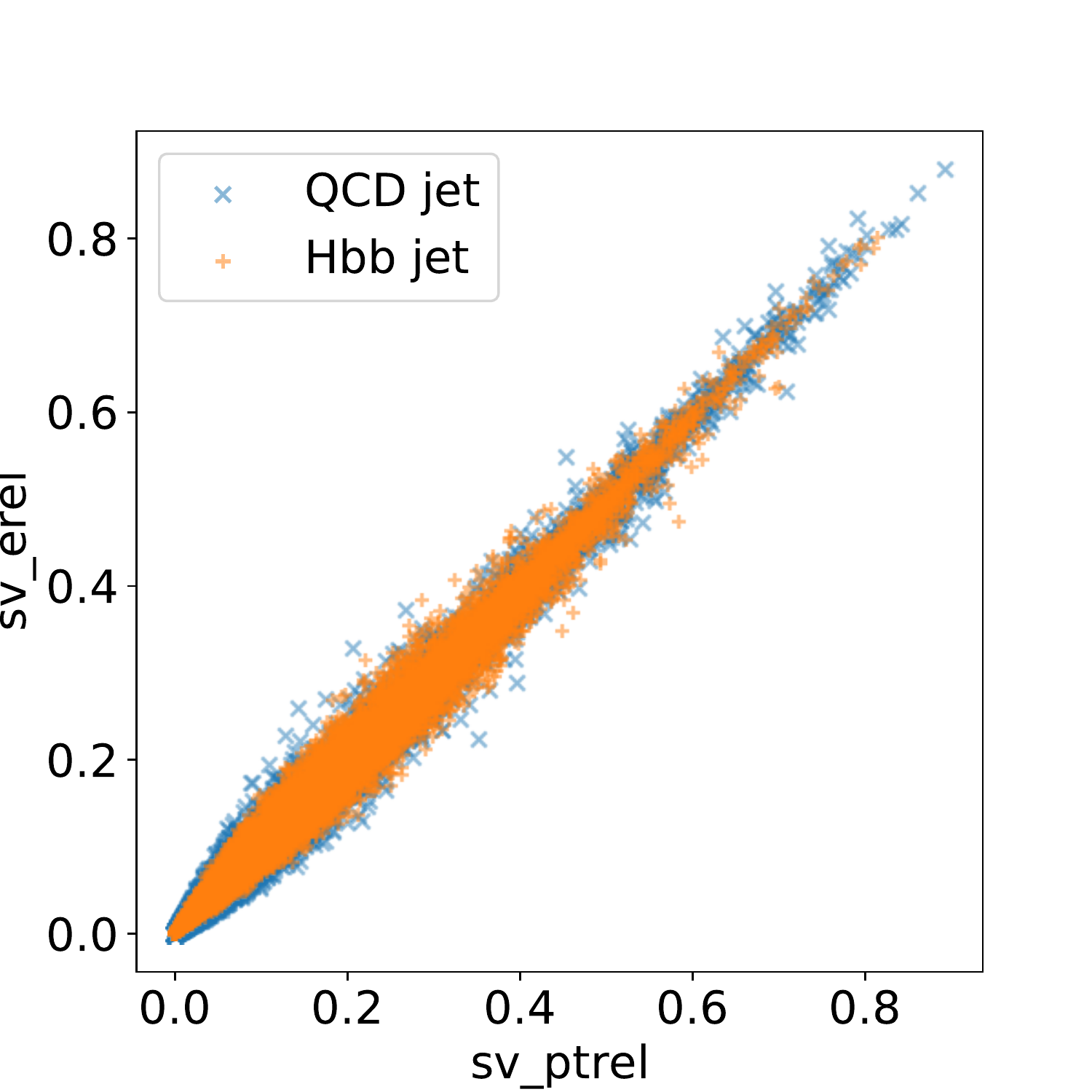}
\label{fig:sverel}
}
\subfloat[]{
\includegraphics[width=0.31\textwidth]{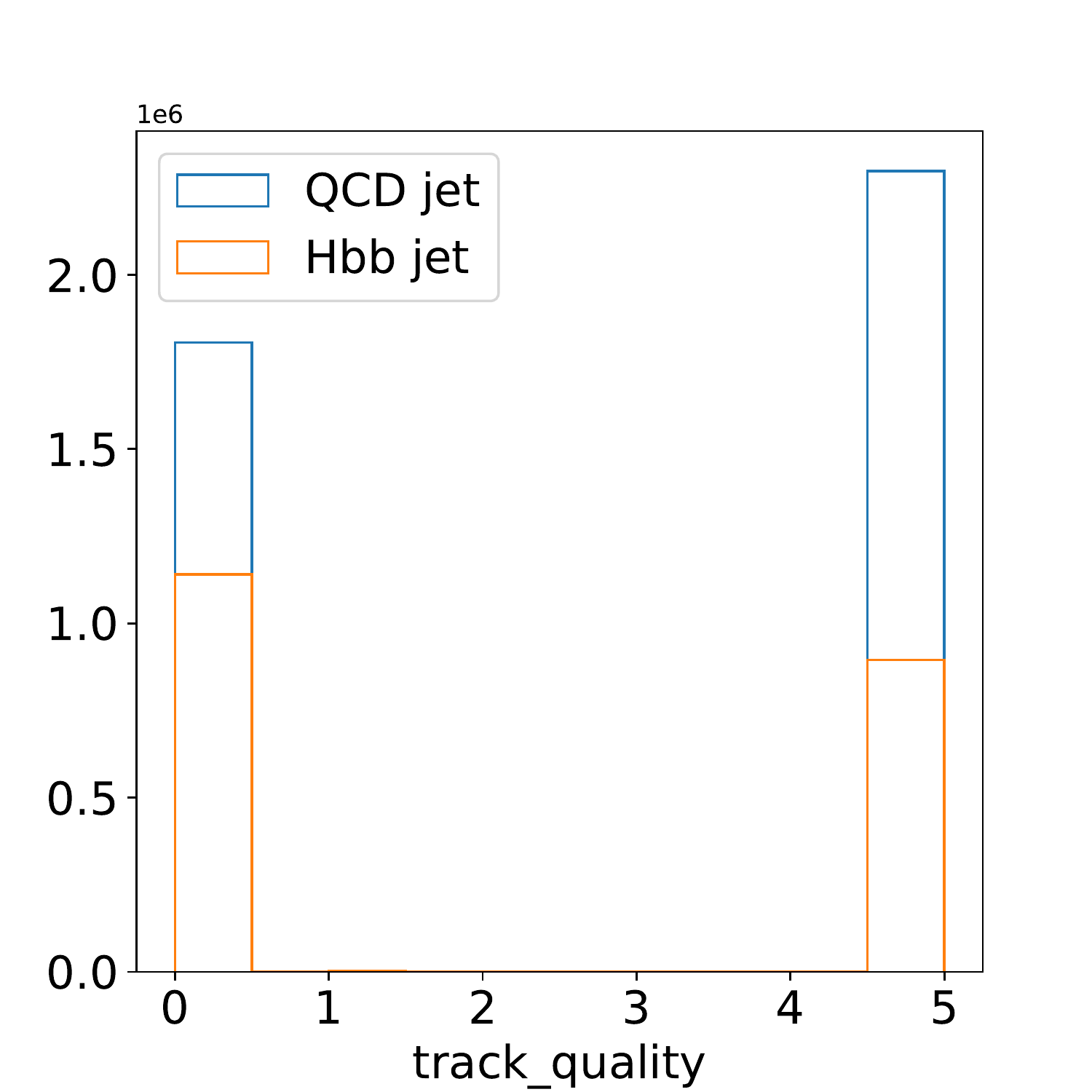}
\label{fig:trackQ}
}
\caption{Scatter plot of \protect\subref{fig:svptrel} \texttt{sv\_ptrel} and \texttt{sv\_pt} and \protect\subref{fig:sverel} \texttt{sv\_ptrel} and \texttt{sv\_erel}. 
\protect\subref{fig:trackQ} shows the distribution of the categorical variable \texttt{track\_quality}. 
}
\label{fig:notrackQ}
\end{figure}

However, the approaches also show some inconsistencies among themselves. The secondary vertex features \texttt{sv\_ptrel} and \texttt{sv\_erel} are assigned very low relevance scores while masking them independently cause very large drops in the AUC score. These variables are highly correlated and both display very large correlation (correlation coefficient of 0.85) with \texttt{sv\_pt} (Figures~\ref{fig:svptrel}~and~\ref{fig:sverel}). The LRP-$\gamma$ method skews their relevance distribution and suppresses the LRP scores for those two variables while assigning a larger score to the variable \texttt{sv\_pt}.

We make an additional observation regarding the importance attributed to the feature called \texttt{track\_quality}. This feature is essentially a qualitative tag regarding the track reconstruction status, and for most of the training data, this has almost identical distribution for both jet categories (Figure~\ref{fig:trackQ}). With such an underlying distribution, it is obvious that this variable doesn't contribute to the classifier's ability to tell apart the jet categories. 
However, the large relevance score associated with it indicates that the classifier's  class-predictive output for each class somehow receives a large contribution from its numerical embedding. 
The model that was trained without these variables, along with the 11 (3) track (vertex) features that report a change in AUC of less than $0.01\%$ converged with an AUC score of $99.00\%$, performing as equally well.

\section{Inspecting the Activation Layers and Model Reoptimization}
\label{sec:NAP}

\begin{wrapfigure}{r}{0.55\textwidth}
\begin{center}
{
\includegraphics[width=0.5\textwidth]{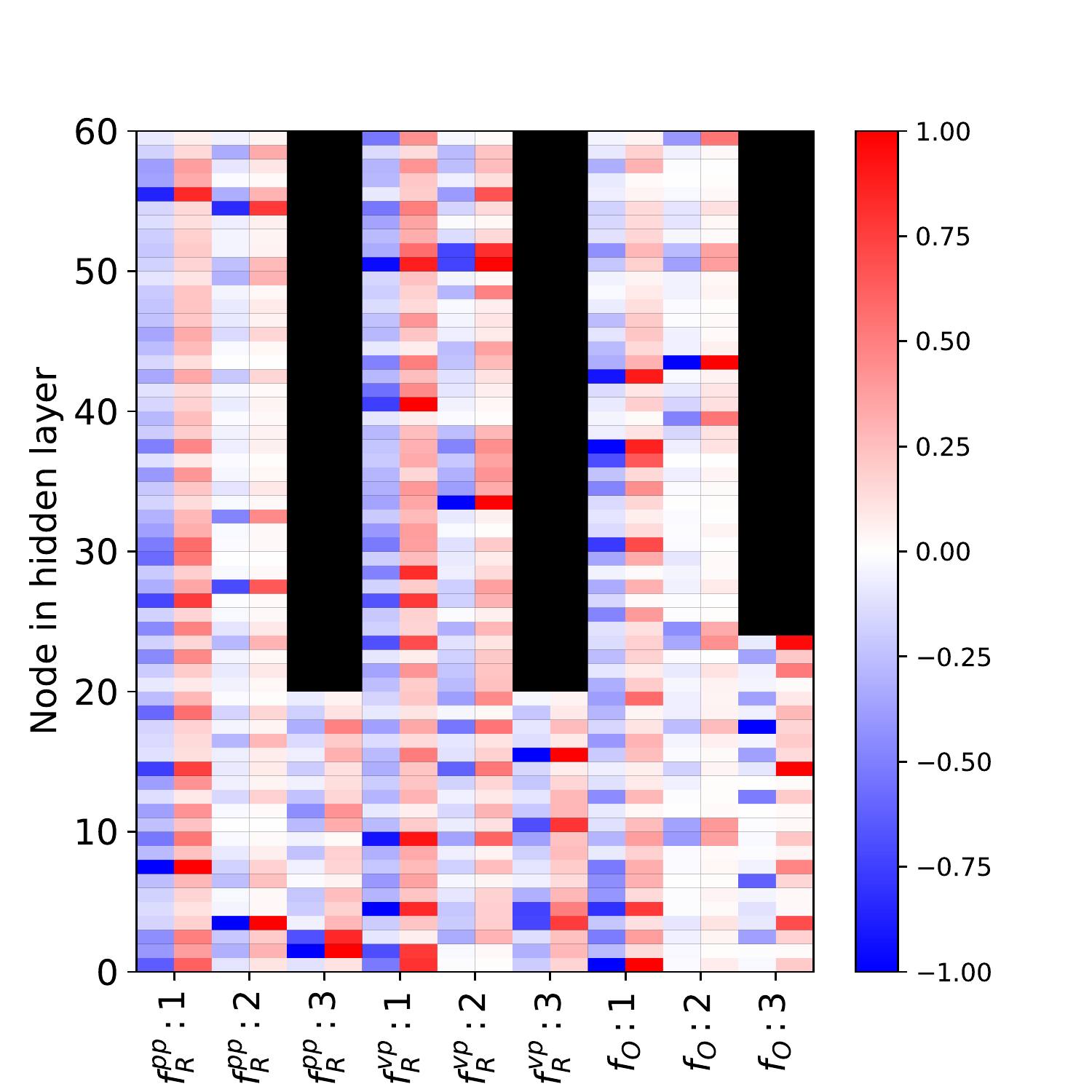}
}
\end{center}
\caption{
2D map of $\mathrm{RNA}$ score for different nodes of the activation layers. To simultaneously visualize the scores for QCD and $H \rightarrow b\bar{b}$ jets, we project the $\mathrm{RNA}$ scores of the former as negative values}
\label{fig:NAP-baseline}
\end{wrapfigure}

As the IN processes the input, it is passed through three different MLPs that approximate arbitrary non-linear functions identifies as $f_R^{pp}, f_R^{vp},$ and $f_O$. 
In order to explore the activity of each neuron and compare it with the activity of neurons in the same layer, 
we define the quantity Relative Neural Activity (RNA)~\cite{khot2022detailed} as $\mathrm{RNA}(j,k;\mathcal{S}) = \frac{\sum_{i=1}^{N} a_{j,k} (s_i)}{\max_j\sum_{i=1}^{N} a_{j,k}(s_i)}$ where $\mathcal{S} =\{s_i\}$ represents a set of samples over which the $\mathrm{RNA}$ score is evaluated. The quantity $a_{j,k} (s_i)$ is the activation of $j$-th neuron in the $k$-th layer when the input to the network is $s_i$. 
Figure~\ref{fig:NAP-baseline} shows the neural activation pattern (NAP) diagram for the baseline model, showing the $\mathrm{RNA}$ scores for the different activation layers. The scores are separately evaluated for QCD and $H \rightarrow b\bar{b}$ categories. To simultaneously visualize these scores, we project the $\mathrm{RNA}$ scores of the former as negative values. The NAP diagram clearly shows that the network's activity level is quite sparse-
while some nodes are playing very important roles in propagating the necessary information, other nodes don't participate as much. We additionally observe that the right until the very last layer of the aggregator network $f_O$, the same nodes show the largest activity level for both jet categories. 
However, different nodes are activated in the last layer for the two jet categories, indicating an effective disentanglement of the jet category information in this layer.

The sparsity of NAP diagram and low feature importance for a number of input features for the baseline IN model indicates that the model can be made simpler, by both reducing the number of input features it relies on and the number of trainable parameters. 
To demonstrate this, we train some alternate variants of the IN models where we drop \texttt{track\_quality}, \texttt{sv\_ptrel}, \texttt{sv\_erel} along with the 11 (3) track (vertex) features that report a change in AUC of less than $0.01\%$.
The details and performance metrics of these models are given in Table~\ref{tab:reopt-models}. 
They demonstrate that the baseline model can be made much simpler without compromising the quality of its performance. As can be seen from the results in Table~\ref{tab:reopt-models}, the ROC-AUC score of the alternate models are very close to that of the baseline model, though the number of trainable parameters is significantly lower.

\begin{table}[!h]
\centering
\begin{tabular}[b]{ |c|c|c|c|c|c| }
  \hline
  $\Delta P$, $\Delta S$ & $h, D_e, D_o$ & Parameters & AUC score &  Sparsity \\
  \hline \hline
  0, 0 (baseline) & 60, 20, 24 & 25554 & 99.02\%  & 0.56 \\
  \hline
   & 32, 16, 16 & 8498 & 98.87\% &   0.52\\
  12, 5 & 32, 8, 8 & 7178 & 98.84\%  &  0.48 \\
   & 16, 8, 8 & 2842 & 98.62\%  & 0.40 \\
  \hline
\end{tabular}
\caption{The performance of the baseline and alternate retrained models with modified hyperparameters. Sparsity is measured by the fraction of activation nodes with an RNA score less than 0.2}
\label{tab:reopt-models}
\end{table}
\vspace{-10pt}

\section{Conclusion}
In this paper we have demonstrated how the application of xAI methods aided with physical intuitions can help identify important features for the task of identifying $H\to b\bar{b} jets$. We additionally propose a novel metric, the RNA score, and an associated visualization tool, the NAP diagram, to investigate information propagation through a model. These tools help understand the sparsity of information propagation and hence optimize model complexity without degrading the model's performance. 

\section*{Acknowledgements}
This work was supported by the U.S. Department of Energy (U.S. DOE), Office of Science, High Energy Physics, under contract number DE-SC0023365 and the FAIR Data program of the U.S. DOE, Office of Science, Advanced Scientific Computing Research, under contract number DE-SC0021258.

\bibliography{PROC-ICHEP-Interp}
\bibliographystyle{JHEP}

\end{document}